\documentclass[oneeqnum,onethmnum,onefignum,onetabnum]{siamltex}

\usepackage{graphicx}
\usepackage{amsfonts}
\usepackage{amssymb}

\newcommand{\Dfp}{D\!f^p}
\newtheorem{conjecture}{\sc Conjecture}
\newtheorem{observation}{\sc Observation}

\title{Efficient Detection of Periodic Orbits in Chaotic Systems by
       Stabilising Transformations
  \thanks{This work was supported by the EPSRC grant GR/S98986/01}}
\author{Jonathan J. Crofts and Ruslan L. Davidchack
  \thanks{Department of Mathematics, University of Leicester,
          Leicester LE1 7RH, UK}}

\begin{document}

\maketitle

\begin{abstract}
An algorithm for detecting periodic orbits in chaotic systems
[Phys. Rev. E, 60 (1999), pp.~6172--6175], 
which combines the set of stabilising transformations proposed by
Schmelcher and Diakonos
[Phys. Rev. Lett., 78 (1997), pp.~4733--4736] 
with a modified semi-implicit Euler iterative scheme and seeding
with periodic orbits of neighbouring periods, has been shown to be
highly efficient when applied to low-dimensional systems. The
difficulty in applying the algorithm to higher-dimensional systems
is mainly due to the fact that the number of the stabilising
transformations grows extremely fast with increasing system
dimension. Here we analyse the properties of stabilising
transformations and propose an alternative approach for constructing
a smaller set of transformations.  The performance of the new
approach is illustrated on the four-dimensional kicked double rotor
map and the six-dimensional system of three coupled H\'{e}non maps.
\end{abstract}



\pagestyle{myheadings}
\thispagestyle{plain}
\markboth{J.~J. CROFTS AND R.~L. DAVIDCHACK}
{DETECTION OF PERIODIC ORBITS}

\section{Introduction}
\label{sec:intro}
Periodic orbits play an important role in the analysis of various
types of dynamical system.  In systems with chaotic behaviour,
unstable periodic orbits (UPOs) form a ``skeleton'' for chaotic
trajectories~\cite{Cvitanovic91}.  A well regarded definition of
chaos~\cite{DevaneyBook} requires the existence of an infinite number
of UPOs that are dense in the chaotic set.  Different geometric and
dynamical properties of chaotic sets, such as natural measure,
Lyapunov exponents, fractal dimensions, entropies~\cite{OttBook},
can be determined from the location and stability properties of the
embedded UPOs.  Periodic orbits are central to the understanding of
quantum-mechanical properties of nonseparable systems: the energy
level density of such systems can be expressed in a semiclassical
approximation as a sum over the UPOs of the corresponding classical
system~\cite{GutzwillerBook}.  Topological description of a chaotic
attractor also benefits from the knowledge of periodic orbits.  For
example, a large set of periodic orbits is highly constraining to the
symbolic dynamics and can be used to extract the location of a
generating partition~\cite{Davidchack00a,Plumecoq00a}.  The
significance of periodic orbits for the experimental study of
dynamical systems has been demonstrated in a wide variety of
systems~\cite{Lathrop89}, especially for the purpose of controlling
chaotic dynamics~\cite{Ott90} with possible application in
communication~\cite{Bollt97}.

It is therefore not surprising that much effort has been put into
the development of methods for locating periodic solutions in
different types of dynamical systems.  In a limited number of cases,
this can be achieved due to the special structure of the systems.
Examples include the Biham-Wenzel method applicable to
H\'{e}non-like maps~\cite{Biham89}, or systems with known and well
ordered symbolic dynamics~\cite{Hansen95}. For generic systems,
however, most methods described in the literature use some type of
an iterative scheme that, given an initial condition (seed),
converges to a periodic orbit of the chaotic system.  In order to
locate all UPOs with a given period $p$, the convergence basin of
each orbit for the chosen iterative scheme must contain at least one
seed.  The seeds are often chosen either at random from within the
region of interest, from a regular grid, or from a chaotic
trajectory with or without close recurrences. Typically, the
iterative scheme is chosen from one of the ``globally'' convergent
methods of quasi-Newton or secant type. However, experience suggests
that even the most sophisticated methods of this type suffer from a
common problem: with increasing period, the basin size of the UPOs
becomes so small that placing a seed within the basin with one of
the above listed seeding schemes is practically
impossible~\cite{Miller00}.

A different approach, which appears to effectively deal with
the problem of reduced basin sizes has been proposed by Schmelcher and
Diakonos (SD)~\cite{Schmelcher97,Schmelcher98}.  The basic idea is to
transform the dynamical system in such a way that the UPOs of the
original system become stable and can be located by simply following
the evolution of the transformed dynamical system.  That is, to locate
period-$p$ orbits of a discrete dynamical system
\begin{equation}
  U\!\!:\quad x_{j+1} = f(x_j),\quad
  f\!\!: {\mathbb R}^n \mapsto {\mathbb R}^n\;,
\label{eq:mapf}\end{equation}
one considers an associated flow
\begin{equation}
  \Sigma\!\!: \quad \frac{dx}{ds} = C g(x)\,,
\label{eq:sflow} \end{equation}
where $g(x) = f^p(x) - x$ and $C$ is an $n\times n$ constant orthogonal
matrix.  It is easy to see that map $f^p(x)$ and flow $\Sigma$
have identical sets of fixed points for any $C$, while $C$ can be
chosen such that unstable period-$p$ orbits of $U$ become stable fixed
points of $\Sigma$.

Since it is not generally possible to choose a single matrix $C$
that would stabilise all UPOs of $U$, the idea
is to find the smallest possible set of matrices
${\mathcal C} = \{C_k\}_{k=1}^K$, such that, for each UPO of $U$,
there is at least one matrix $C \in {\mathcal C}$ that
transforms the unstable orbit of $U$ into a stable fixed point
of $\Sigma$.  To this end, Schmelcher and Diakonos have put forward
the following conjecture~\cite{Schmelcher97}:
\begin{conjecture}\label{conj:sd}
Let ${\mathcal C}_{\mathrm{SD}}$ be the set of all $n\times n$
orthogonal matrices with only $\pm 1$ non-zero entries. Then, for
any $n\times n$ non-singular real matrix $G$, there exists a matrix
$C \in {\mathcal C}_{\mathrm{SD}}$ such that all eigenvalues of the
product $CG$ have negative real parts.
\end{conjecture}
\begin{observation}The set ${\mathcal C}_{\mathrm{SD}}$ forms a
group isomorphic to the Weyl group $B_n$~\cite{HumphreysBook}, i.e.
the symmetry group of an $n$-dimensional hypercube.  The number of
matrices in ${\mathcal C}_{\mathrm{SD}}$ is $K = 2^n n!$.
\end{observation}

The above conjecture has been verified for $n \le 2$ and appears to be
true for $n > 2$, but, thus far, no proof has been presented.
According to this conjecture, any periodic orbit, whose stability matrix
does not have eigenvalues equal to one, can be transformed into a
stable fixed point of $\Sigma$ with $C \in {\mathcal C}_{\mathrm{SD}}$.
In practice, to locate periodic orbits of the map $U$, we try to
integrate the flow $\Sigma$ from a given initial condition (seed) using
different matrices from the set ${\mathcal C}_{\mathrm{SD}}$.
Some of the resulting trajectories will converge to fixed points,
while others will fail to do so, either leaving the region of interest
or failing to converge within a specified number of steps.

The main advantage of the SD approach is that the convergence basins
of the stabilised UPOs appear to be much larger than the basins
produced by other iterative
schemes~\cite{Schmelcher98,Klebanoff01,Davidchack01b},
making it much easier to select a useful seed.  Moreover, depending
on the choice of the stabilising transformation, the SD method may
converge to several different UPOs from the same seed.

The flow $\Sigma$ can be integrated by any off-the-shelf numerical
integrator.  Schmelcher and Diakonos have enjoyed considerable
success using a simple Euler method.  However, the choice of the
integrator for this problem is governed by considerations very
different from those typically used to construct an ODE solver.
Indeed, to locate a fixed point of the flow, it may not be very
efficient to follow the flow with some prescribed accuracy.
Therefore, local error considerations, for example, are not as
important.  Instead, the goal is to have a solver that can reach the
fixed point in as few integration steps as possible.  In fact, as
shown by Davidchack and Lai~\cite{Davidchack99c}, the efficiency of
the method can be improved dramatically when the solver is
constructed specifically with the above goal in mind. In particular,
recognizing the typical stiffness of the flow $\Sigma$, Davidchack
and Lai have proposed a modified semi-implicit Euler method:
\begin{equation}
 x_{j+1} = x_j + [\beta s_j C^{\mathsf T} - G_j]^{-1}g(x_j)\;,
\label{eq:itrDL} \end{equation}
where $\beta > 0$ is a scalar parameter, $s_j \equiv ||g(x_j)||$
is an $L_2$ norm, $G_j \equiv Dg(x_j)$ is the Jacobian matrix, and
``${\mathsf T}$'' denotes transpose.  Note that, away from the root
of $g$, the above iterative scheme is a semi-implicit Euler method
with step size $h = (\beta s_j)^{-1}$ and, therefore, can follow the
flow $\Sigma$ with a much larger step size than an explicit integrator
(e.g. Euler or Runge-Kutta).  Close to the root, the proposed scheme
can be shown to converge quadratically~\cite{Klebanoff01}, analogous
to the Newton-Raphson method.

Another important ingredient of the algorithm presented
in~\cite{Davidchack99c} is the seeding with already detected
periodic orbits of neighbouring periods.  This seeding scheme
appears to be superior to the typically employed schemes and enables
fast detection of plausibly all\footnote{It is not possible to
prove, within our approach, the completeness of the detected sets of
UPOs. Rather, our assertion of completeness is based on the
plausibility argument. The following three criteria are used for the
validation of the argument:\\
i) Methods based on rigorous numerics (e.g. in~\cite{Galias01}) have
located the same UPOs in cases where such comparison is possible
(usually for low periods, since these methods are less
efficient).\\
ii) Our search strategy scales with the period $p$ (see
\S\ref{sec:numres} and~\cite{Davidchack01b}).  If we can tune it to
locate all UPOs for low periods (where we can verify the
completeness using (i)), it is likely (but not provably) capable
of locating all UPOs of higher periods as well.\\
iii) For maps with symmetries, we test the completeness by verifying
that all the symmetric partners for all detected UPOs have been
found (see \S\ref{sec:drm} and \S\ref{sec:chm}).} periodic orbits of
increasingly larger periods in generic low-dimensional chaotic
systems. For example, for the Ikeda map at traditional parameter
values, the algorithm presented in~\cite{Davidchack99c} was able to
locate plausibly all periodic orbits up to period 22 for a total of
over $10^6$ orbit points. Obtaining a comparable result with
generally employed techniques requires an estimated $10^5$ larger
computational effort.

While the stabilisation approach is straightforward for relatively
low-dimensional systems, direct application to higher-dimensional
systems is much less efficient due to the rapid growth of the number
of matrices in ${\mathcal C}_{\mathrm{SD}}$. Even though it appears
that, in practice, far fewer transformations are required to
stabilise all periodic orbits of a given chaotic
system~\cite{Pingel04}, the sufficient subset of transformations is
not known a priori.  It is also clear that the route of constructing
a universal set of transformations is unlikely to yield substantial
reduction in the number of such transformations. For instance, a
smaller set of universal transformations with $K = (n+1)!$, which is
isomorphic to the Weyl group $A_n$, is sufficient to stabilise all
types of periodic orbits for $n < 4$, but can be shown to fail for
certain types of orbits when $n \ge 4$.  Therefore, a more promising
way of using stabilising transformations for locating periodic
orbits in high-dimensional systems is to design such transformations
based on the information about the properties of the system under
investigation.

In this Article, we propose to construct stabilising transformations
based on the knowledge of the stability matrices of already detected
periodic orbits (used as seeds).  The advantage of our approach is
in a substantial reduction of the number of transformations, which
increases the efficiency of the detection algorithm, especially in
the case of higher-dimensional systems. The layout of the paper is
as follows. In \S\ref{sec:stab2d} we study the properties of the
stabilising transformations for $n = 2$ and their relationship to
the properties of the stability matrix of a periodic orbit.  In
\S\ref{sec:ext} we extend the analysis to higher-dimensional systems
and show how to construct stabilising transformations using the
knowledge of the stability matrices of already detected periodic
orbit points. In particular, we argue that the stabilising
transformations depend essentially on the signs of unstable
eigenvalues and the directions of the corresponding eigenvectors of
the stability matrices. Section \ref{sec:numres} illustrates the
application of the new stabilising transformations to the detection
of periodic orbits in a four-dimensional kicked double rotor map and
a six-dimensional system of three coupled H\'{e}non maps.  We
conclude with the summary and discussion of possible further
developments of the stabilising transformations approach in
\S\ref{sec:last}.

\section{Stabilising transformations in two dimensions}
\label{sec:stab2d}
The stability of a fixed point $x^\ast$ of the flow $\Sigma$ is
determined by the real parts of the eigenvalues of the matrix $C G$,
where $G \equiv Dg(x^\ast)$ is the Jacobian matrix of $g(x)$ evaluated
at $x^\ast$.  For $x^\ast$ to be a stable fixed point of
$\Sigma$, the matrix $C$ has to be such that all the eigenvalues of
$CG$ have negative real parts.  In order to understand what
properties of $G$ determine the choice of a particular stabilising
transformation $C$, we use the following parametrisation for
the general two-dimensional orthogonal matrices:
\begin{equation}
C_{s,\alpha} = \left(\!\!\begin{array}{cc}
s \cos \alpha & \sin \alpha \\
-s \sin \alpha & \cos \alpha \end{array} \!\!\right)
\label{eq:par2d} \end{equation}
where $s = \pm 1$ and $-\pi < \alpha \le \pi$.
When $\alpha = -\pi/2,~0,~\pi/2$, or $\pi$, we
obtain the set of matrices ${\mathcal C}_{\mathrm{SD}}$.
For example, $C_{1,\pi/2} = {\hspace*{4pt}0\hspace{3.5pt}1 \choose
-1\hspace{2pt}0}$ and $C_{-1,\pi} =
{\!1\hspace*{5pt}0 \choose \hspace*{2pt}0\hspace*{2pt}-1}$.

If we write $G \equiv g_{ij}$, $(i,j = 1,2)$, then the eigenvalues
of $C_{s,\alpha}G$ are given by the following equations:
\begin{equation}
  \sigma_{1,2} = -A\cos(\alpha - \theta) \pm
  \sqrt{A^2\cos^2(\alpha - \theta) - s \det G}
\label{eq:eigs1} \end{equation}
where $\det G = g_{11} g_{22} - g_{12} g_{21}$,
$~A = \frac{1}{2}\sqrt{(s g_{11} + g_{22})^2
    + (s g_{12} - g_{21})^2}$, and
\begin{equation}\label{eq:theta}
  \tan\theta = \frac{s g_{12} - g_{21}}{-s g_{11} - g_{22}}\:,
  \qquad -\pi < \theta \le \pi\,.
\end{equation}
Note that the signs of the numerator and denominator are significant
for defining angle $\theta$ in the specified range and should not be
canceled out.  It is clear from Eq.(\ref{eq:eigs1}) that both
eigenvalues have negative real parts when
\begin{equation}\label{eq:cond}
  s = \bar{s} \equiv {\mathrm{sgn}}\,\det G,\qquad
  \mbox{and} \qquad | \alpha - \theta | < \textstyle\frac{\pi}{2}\,,
\end{equation}
provided that $\det G \neq 0$.  This result proves the validity of
Conjecture \ref{conj:sd} for $n = 2$.  Moreover, it shows that there
are typically two matrices in ${\mathcal C}_{\mathrm{SD}}$ that
stabilise a given fixed point.

Parameter $\theta$ clearly plays an important role in the above
analysis.  The following theorems show its relationship to the
eigenvalues and eigenvectors of the stability matrix of a fixed point.
\begin{theorem}
Let $x^\ast$ be a saddle fixed point of $f^p(x): {\mathbb R}^2 \mapsto
{\mathbb R}^2$ whose stability matrix $\Dfp(x^\ast)$ has eigenvalues
$\lambda_{1,2}$ such that $|\lambda_2| < 1 < |\lambda_1|$ and
eigenvectors defined by the polar angles $0 \le \phi_{1,2} < \pi$,
i.e. $v_{1,2} = (\cos\phi_{1,2}, \sin\phi_{1,2})^{\mathsf T}$.
Then the following is true for angle $\theta$ defined in
Eq.~(\ref{eq:theta}):\\
\underline{Case 1: $\lambda_1 < -1$}
\begin{equation}\label{eq:th1case1a}
  \textstyle \theta \in \left(-\frac{\pi}{2}, \frac{\pi}{2}\right)\,.
\end{equation}
Moreover, if $|\lambda_1| \gg 1$, then
\begin{equation}
  \theta \approx (\phi_1 - \phi_2) (\mathrm{mod~}\pi) -
  \textstyle \frac{\pi}{2}\,.
\label{eq:th1case1b} \end{equation}
\underline{Case 2: $\lambda_1 > 1$}
\begin{equation}\label{eq:th1case2}
  \theta = \left\{\begin{array}{cc}
  \frac{3\pi}{2} - \phi_1 - \phi_2\,, & 0 < \phi_1 - \phi_2 < \pi\,,\\
  \frac{\pi}{2} - \phi_1 - \phi_2\,, & -\pi < \phi_1 - \phi_2 < 0\,.\\
  \end{array} \right.
\end{equation}
\end{theorem}
\begin{proof}
Matrix $G = \Dfp(x^\ast) - I$, where $I$ is the identity matrix, can
be written as follows:
\begin{equation}\label{eq:eigdec}
  G \equiv \left(\!\!\begin{array}{cc} g_{11} \!&\! g_{12} \\
  g_{21} \!&\! g_{22}\end{array} \!\!\right) =
  \left(\!\!\begin{array}{cc} \cos\phi_1 \!&\! \cos\phi_2 \\
  \sin\phi_1 \!&\! \sin\phi_2\end{array} \!\!\right)
  \left(\!\!\begin{array}{cc} \lambda _{1}-1 \!&\! 0 \\
  0 \!&\! \lambda _{2}-1\end{array} \!\!\right)
  \left(\!\!\begin{array}{cc} \cos\phi_1 \!&\! \cos\phi_2 \\
  \sin\phi_1 \!&\! \sin\phi_2\end{array} \!\!\right)^{-1}
\end{equation}
\underline{Case 1}:  Since
$\det G = (\lambda_1 - 1)(\lambda_2 - 1) > 0$ we set $s = 1$ and
obtain from Eq.~(\ref{eq:theta}):
\begin{equation}
  \tan\theta = \frac{(\lambda_1-\lambda_2)\cot(\phi_1-\phi_2)}
  {2-\lambda_1-\lambda_2}\,,
\label{eq:theta_c1} \end{equation} where, just like in
Eq.~(\ref{eq:theta}), as well as in Eqs.~(\ref{eq:theta_c2}) and
(\ref{eq:theta_t1}) below, the signs of the numerator and
denominator should not be canceled out. Since $2-\lambda_1-\lambda_2
> 0$, we have that $\cos\theta > 0$ or
\[  \textstyle\theta\in\left(-\frac{\pi}{2}, \frac{\pi}{2}\right)\,. \]
For $|\lambda_1| \gg 1$, Eq.~(\ref{eq:theta_c1}) yields:
\[  \tan\theta \approx -\cot(\phi_1-\phi_2)\,, \]
and, given Eq.~(\ref{eq:th1case1a}), the result in
Eq.~(\ref{eq:th1case1b}) immediately follows.\newline
\underline{Case 2}: In this case
$\det G = (\lambda_1 - 1)(\lambda_2 - 1) < 0$, so,
from Eq.~(\ref{eq:theta}) with $s = -1$:
\begin{eqnarray}\label{eq:theta_c2}
  \tan\theta &=& \frac{(\lambda_2-\lambda_1)\cos(\phi_1+\phi_2)/
  \sin(\phi_1-\phi_2)}{(\lambda_2-\lambda_1)\sin(\phi_1+\phi_2)/
  \sin(\phi_1-\phi_2)}\\ &=& \frac{-\cos(\phi_1+\phi_2)/
  \sin(\phi_1-\phi_2)}{-\sin(\phi_1+\phi_2)/\sin(\phi_1-\phi_2)}\,,
\nonumber
\end{eqnarray}
since $\lambda_2-\lambda_1 < 0$. The result in
Eq.~(\ref{eq:th1case2}) follows.\qquad\end{proof}
\begin{theorem}
Let $x^\ast$ be a spiral fixed point of $f^p(x): {\mathbb R}^2 \mapsto
{\mathbb R}^2$ whose stability matrix $\Dfp(x^\ast)$ has eigenvalues
$\lambda_{1,2} = \lambda \pm \mathrm{i}\omega$.  Then
\begin{eqnarray}\label{eq:th2}
  \theta & \in & \textstyle \left(-\frac{\pi}{2}, \frac{\pi}{2}\right)
  \hspace{1.9cm}\mathrm{if}\quad \lambda < 1\,,\\
  \theta & \in & \textstyle \left(-\pi, -\frac{\pi}{2}\right)\cup
  \left(\frac{\pi}{2}, \pi\right)
  \quad\mathrm{if}\quad \lambda > 1\,.\nonumber
\end{eqnarray}
\end{theorem}
\begin{proof}
The stability matrix can be decomposed as follows:
\begin{equation}
  \Dfp(x^\ast)=\left(\!\!\begin{array}{cc}
  \cos\phi & \mathrm{e}^\eta \\
  \sin\phi & 0\end{array} \!\!\right)
  \left(\!\!\begin{array}{cc} \lambda & \omega \\
  -\omega & \lambda \end{array} \!\!\right)
  \left(\!\!\begin{array}{cc} \cos\phi & \mathrm{e}^\eta \\
  \sin\phi & 0\end{array} \!\!\right)^{-1}\,,
\end{equation}
where $\eta \in \mathbb{R}$.  Given that $G = \Dfp(x^\ast) - I$, we
have from Eq.~(\ref{eq:theta}):
\begin{equation}\label{eq:theta_t1}
  \tan\theta = \frac{-\omega\cosh\eta/\sin\phi}{1-\lambda}\,.
\end{equation}
The result in Eq.~(\ref{eq:th2}) follows from the sign of the
denominator.\qquad\end{proof}

The key message of the above theorems is that the stabilising
transformation matrix depends mostly on the directions of the
eigenvectors and the signs of the unstable\footnote{That is,
eigenvalues whose magnitude is larger than one.}
eigenvalues (or their real parts), and only marginally on the actual
magnitudes of the eigenvalues.  This means that a transformation that
stabilises a given fixed point $x^\ast$ of $f^p$ will also stabilise
fixed points of all periods with similar directions of eigenvectors
and signs of the unstable eigenvalues.  In the next Section, we will
show how this observation can be used to construct stabilising
transformations for efficient detection of periodic orbits in
systems with $n > 2$.

\section{Extension to higher-dimensional systems}
\label{sec:ext}
To extend the analysis of the preceding Section to higher-dimensional
systems, we note that the matrix $C_{\bar{s},\theta}$, as defined
by Eqs.~(\ref{eq:par2d}), (\ref{eq:theta}),  and (\ref{eq:cond}), is
closely related to the orthogonal part of the {\em polar
decomposition} of $G$~\cite{HalmosBook}.  Recall that any
non-singular $n\times n$ matrix can be uniquely represented
as a product
\begin{equation}\label{eq:polar}
  G = QB\,,
\end{equation}
where $Q$ is an orthogonal matrix and $B$ is a symmetric positive
definite matrix.  The following theorem provides the link between
$C_{\bar{s},\theta}$ and $Q$ for $n = 2$:
\begin{theorem}
Let $G \in \mathbb{R}^{2\times 2}$ be a non-singular matrix
with the polar decomposition $G = QB$, where $Q$ is an orthogonal
matrix and $B$ is a symmetric positive definite matrix.
Then matrix $C_{\bar{s},\theta}$, as defined by Eqs.~(\ref{eq:par2d}),
(\ref{eq:theta}) and (\ref{eq:cond}), is related to $Q$ as follows:
\begin{equation}\label{eq:qt1}
  C_{\bar{s},\theta} = -Q^{\mathsf T}
\end{equation}
\end{theorem}
\begin{proof}
Since $C_{\bar{s}, \theta}$ is an orthogonal matrix by
definition, it is sufficient to prove that $C_{\bar{s}, \theta}G$ is
symmetric negative definite.  Then, by the uniqueness of the polar
decomposition, it must be equal to $-B$.

Denote by $b_{ij}$ the element in the $i$-th row and $j$-th column
of $C_{\bar{s}, \theta}G$.  We must show that $b_{12} = b_{21}$.
Using Eq.~(\ref{eq:theta}), we have that
\begin{eqnarray}
  b_{12} &=& \bar{s}g_{12}\cos\theta + g_{22}\sin\theta\\
  &=& \left[\bar{s}g_{12} + g_{22}\frac{\bar{s}g_{12}-g_{21}}
  {-\bar{s}g_{11}-g_{22}}\right]\cos\theta \nonumber \\
  &=& \left[\frac{g_{11}g_{12}+g_{21}g_{22}}
  {\bar{s}g_{11}+g_{22}}\right]\cos\theta\,, \nonumber
\end{eqnarray}
and similarly
\begin{eqnarray}
  b_{21} &=& g_{21}\cos\theta - \bar{s}g_{11}\sin\theta\\
  &=& \left[g_{21} - \bar{s}g_{11}\frac{\bar{s}g_{12} -
  g_{21}}{-\bar{s}g_{11} - g_{22}}\right]\cos\theta\nonumber \\
  &=& \left[\frac{g_{11}g_{12}+g_{21}g_{22}}
  {\bar{s}g_{11}+g_{22}}\right]\cos\theta\,, \nonumber
\end{eqnarray}
hence the matrix $C_{\bar{s}, \theta}G$ is symmetric.  Since, by
definition, $\theta$ and $\bar{s}$ are chosen such that the
eigenvalues of $C_{\bar{s}, \theta}G$ are negative, the matrix
$C_{\bar{s}, \theta}G$ is negative definite.  Finally,
by the uniqueness of the polar decomposition,
\[ C_{\bar{s}, \theta}G = -B = -Q^{\mathsf T}G\,, \]
which completes the proof.\qquad\end{proof}

For $n > 2$, we can always use the polar decomposition to construct
a transformation that will stabilise a given fixed point.
Indeed, if a fixed point $x^\ast$ of an $n$-dimensional flow has
a non-singular matrix $G \equiv Dg(x^\ast)$, then we can
calculate the polar decomposition $G = QB$ and use
\begin{equation}\label{eq:qt2}
  C = -Q^{\mathsf T}\;,
\end{equation}
to stabilise $x^\ast$. Moreover, by analogy with the two-dimensional
case, we can expect that the same matrix $C$ will also stabilise
fixed points $\tilde{x}$ with the matrix $\tilde{G} \equiv
Dg(\tilde{x})$, as long as the orthogonal part $\tilde{Q}$ of the
polar decomposition $\tilde{G} = \tilde{Q}\tilde{B}$ is sufficiently
close to $Q$. More precisely,
\begin{observation} \label{obs:stab}
$C$ will stabilise $\tilde{x}$, if all eigenvalues of the product
$Q\tilde{Q}^\mathsf{T}$ have positive real parts.
\end{observation}

We base this observation on the following corollary of Lyapunov's
stability theorem~\cite{HornBook2}:
\begin{corollary}
Let $B \in \mathbb{R}^{n\times n}$ be a positive definite symmetric
matrix.  If $Q \in \mathbb{R}^{n\times n}$ is an orthogonal matrix
such that all its eigenvalues have positive real parts, then all the
eigenvalues of the product $QB$ have positive real parts as well.
\end{corollary}
\begin{proof}
According to Lyapunov's theorem, a matrix $A \in \mathbb{R}^{n\times
n}$ has all eigenvalues with positive real parts if and only if
there exists a symmetric positive definite $G \in
\mathbb{R}^{n\times n}$ such that $GA + A^\mathsf{T}G = H$ is
positive definite.

Let $A = QB$ and let's choose $G$ in the form $G =\frac{1}{2}
QB^{-1}Q^\mathsf{T}$.  Since $B$ is positive definite, its inverse
$B^{-1}$ is also positive definite, and, since $G$ and $B^{-1}$ are
related by a congruence transformation, according to Sylvester's
inertia law~\cite{HornBook1}, $G$ is also positive definite.  Now,
\[ GQB + (QB)^\mathsf{T}G =
\textstyle\frac{1}{2}QB^{-1}Q^\mathsf{T}QB +
\frac{1}{2}BQ^\mathsf{T}QB^{-1}Q^\mathsf{T} = \frac{1}{2}[Q +
Q^\mathsf{T}]\;.\] Therefore, $QB$ has eigenvalues with positive
real parts if and only if $\frac{1}{2}[Q + Q^\mathsf{T}]$ is
positive definite.  The proof is completed by observing that, for
orthogonal matrices, the eigenvalues of $\frac{1}{2}[Q +
Q^\mathsf{T}]$ are equal to the real parts of the eigenvalues of
$Q$.\qquad\end{proof}

Note that Observation \ref{obs:stab} is a direct generalisation of
conditions in Eq.~(\ref{eq:cond}) which are equivalent to requiring
that the eigenvalues of $C_{s,\alpha}C_{\bar{s}, \theta}^\mathsf{T}$
have positive real parts.

In the scheme where already detected periodic orbits are used as
seeds to detect other orbits~\cite{Davidchack99c}, we can use $C$ in
Eq.~(\ref{eq:qt2}) as a stabilising matrix for the seed $x^\ast$.
Based on the analysis in \S\ref{sec:stab2d}, this will allow us to
locate a periodic orbit in the neighbourhood of $x^\ast$ with
similar invariant directions and the same signs of the unstable
eigenvalues.  Note, however, that the neighbourhood of the seed
$x^\ast$ can also contain periodic orbits with the similar invariant
directions but with some eigenvalues having the opposite sign
(i.e.~orbits with and without reflections).  To construct
transformations that would stabilise such periodic orbits, we can
determine the eigenvalues and eigenvectors of the stability matrix
of $x^\ast$
\begin{equation}
  \Dfp(x^\ast) = V\Lambda V^{-1}\,,
\label{eq:gev} \end{equation}
where $\Lambda \equiv \mathrm{diag}(\lambda_1,\ldots,\lambda_n)$ is
the diagonal matrix of eigenvalues of $\Dfp(x^\ast)$ and $V$ is
the matrix of eigenvectors, and then calculate the polar decomposition
of the matrix
\begin{equation}
  \hat{G} = V(S\Lambda - I)V^{-1}\,,
\label{eq:gpev} \end{equation}
where $S = \mathrm{diag}(\pm 1, \pm 1,\ldots,\pm 1)$.  Note that,
as follows from the analysis in \S\ref{sec:stab2d} for $n = 2$
and numerical evidence for $n > 2$, changing the sign of a stable
eigenvalue will not result in a substantially different stabilising
transformation.  Therefore, we restrict our attention to the following
subset of $S$:
\begin{equation}\label{eq:signs}
  S_{ii} = \left\{ \begin{array}{rl} \pm 1, & |\lambda_i| > 1\,,\\
  1, & |\lambda_i| < 1\,,   \end{array} \right.
  \quad\mathrm{for}\quad i=1,\ldots,n\,.
\end{equation}
For a seed with $k$ real unstable eigenvalues, this results in $2^k$
possible transformations.  Note that, on the one hand, this set is
much smaller than ${\mathcal C}_{\mathrm{SD}}$, while, on the other
hand, it allows us to target all possible types of periodic orbits
that have invariant directions similar to those at the seed.

\section{Numerical results}\label{sec:numres}
In this Section we illustrate the performance of the new stabilising
transformations on a four-dimensional kicked double rotor
map~\cite{Romeiras92} and a six-dimensional system of three coupled
H\'{e}non maps~\cite{Politi92}.  Both systems are highly chaotic and
the number of UPOs is expected to grow rapidly with increasing
period.  The goal is to locate {\em all} UPOs of increasingly larger
period.  Of course, the completeness of the set of orbits for each
period cannot be guaranteed, but it can be established with high
degree of certainty by using the plausibility criteria outlined in
the Introduction.

In order to start the detection process, we need to have a small set
of periodic orbits (of period $p > 1$) that can be used as seeds.
Such orbits can be located using, for example, random seeds and the
standard Newton-Raphson method (or the scheme in
Eq.~(\ref{eq:itrDL}) with $\beta = 0$).  We can then use these
periodic orbits as seeds to construct the stabilising
transformations and detect more UPOs with higher efficiency.  The
process can be iterated until we find no more orbits of a given
period. In our previous work~\cite{Davidchack99c,Davidchack01b} we
showed that for two-dimensional maps such as H\'{e}non and Ikeda it
is sufficient to use period-$(p-1)$ orbits as seeds to locate
plausibly all period-$p$ orbits.  For higher-dimensional systems,
such as those considered in the present work, these seeds may not be
sufficient. However, it is always possible to use more seeds by, for
example, locating some of the period-$(p+1)$ orbits, which can then
be used as seeds to complete the detection of period-$p$ orbits. The
following recipe can be used as a general guideline for developing a
specific detection scheme for a given system: {\em
\begin{remunerate}
\item Find a set of orbit points of low period using random seeds
and the iterative scheme in Eq.~(\ref{eq:itrDL}) with $\beta = 0$
(i.e. the Newton-Raphson scheme).
\item To locate period-$p$ orbits, first use period-$(p-1)$ orbits
as seeds. For each seed $x_0$, construct $2^{k}$ stabilising
transformations $C$ using Eqs.~(\ref{eq:gev}-\ref{eq:signs}), where
$k$ is the number of unstable eigenvalues of $D\!f^{p-1}(x_0)$.
\item Starting from $x_0$ and with a fixed value of $\beta > 0$
use the iteration scheme in Eq.~(\ref{eq:itrDL}) to construct a
sequence $\{x_{i}\}$ for each of the $2^{k}$ stabilising
transformations. If a sequence converges to a point,
check whether it is a new period-$p$ orbit point, and if so,
proceed to find a complete orbit by iterating the map $f$.
\item Repeat steps $2-3$ for several $\beta$ in order to determine
the optimal value of this parameter (see explanation below).
\item Repeat steps $2-4$ using newly found period-$p$ points
as seeds to search for period-$(p+1)$ orbits.
\item Repeat steps $2-4$ using incomplete set of period-$(p+1)$
orbits as seeds to find any missing period-$p$ orbits.
\end{remunerate} }

Although we know that the action of $\beta$ is to increase the basin
size of the stabilised points, it is not known {\em a priori} what
values of $\beta$ to use for a given system and period.  Monitoring
the fraction of seeds that converge to periodic orbits, we observe
that it grows with increasing $\beta$ until it reaches saturation,
indicating that the iterative scheme faithfully follows the flow
$\Sigma$.  On the other hand, larger $\beta$ translates into smaller
integration steps and, therefore, longer iteration sequences.  Thus
the optimal value of $\beta$ is just before the saturation point. As
demonstrated in our previous work~\cite{Davidchack99c} and observed
in the numerical examples presented in the following sections, this
value appears to scale exponentially with the period and can be
estimated based on the information about the detection pattern at
lower periods.

The stopping criteria in step 3, which we use in the numerical
examples discussed below, are as follows.  The search for UPOs is
conducted within a rectangular region containing a chaotic invariant
set.  The sequence $\{x_{i}\}$ is terminated if (i) $x_{i}$ leaves
the region, (ii) $i$ becomes larger than a pre-defined maximum
number of iterations (we use $i > 100+5\beta$ ), (iii) the sequence
converges, such that $\|g(x_i)\| < \mbox{\em Tol}_g$.  In cases (i)
and (ii) a new sequence is generated from a different seed and/or
with a different stabilising matrix.  In case (iii) five Newton
iterations are applied to $x_i$ to allow convergence to a fixed
point to within the round-off error.  A point $x^\ast$ for which
$\|g(x^\ast)\|$ is the smallest is identified with a fixed point of
$f^p$.  The maximum round-off error over the set ${\cal X}_p$ of all
detected period-$p$ orbit points
\begin{equation}\label{eq:maxe}
  \epsilon_\mathrm{max}(p)=\max\{\|g(x^\ast)\|: x^\ast\in{\cal X}_p\}
\end{equation}
is monitored in order to assess the accuracy of the detected orbits.

To check if the newly detected orbit is different from those already
detected, its distance to other orbit points is calculated: if
$\|x^\ast-y^\ast\|_\infty > \mbox{\em Tol}_x$ for all previously
detected orbit points $y^\ast$, then $x^\ast$ is a new orbit point.
Even for a large number of already detected UPOs, this check can be
done very quickly by pre-sorting the detected orbit points along one
of the system coordinates and performing a binary search for the
points within $\mbox{\em Tol}_x$ of $x^\ast$.  The infinity norm in
the above expression is used for the computational efficiency of
this check.

The minimum distance between orbit points
\begin{equation}\label{eq:mind}
    d_\mathrm{min}(p)=\min\{\|x^\ast-y^\ast\|_\infty :
    x^\ast, y^\ast \in{\cal X}_p\}
\end{equation}
is monitored and the algorithm is capable of locating all isolated
UPOs of a given period $p$ as long as $\epsilon_\mathrm{max}(p) <
\mbox{\em Tol}_g \lesssim \mbox{\em Tol}_x < d_\mathrm{min}(p)$.
Since typically $\epsilon_\mathrm{max}(p)$ increases and
$d_\mathrm{min}(p)$ decreases with $p$ (see Tables \ref{tab:drmp}
and \ref{tab:chm}), the above conditions can be satisfied up to some
period, after which higher-precision arithmetics needs to be used in
the evaluation of the map. For the numerical examples presented in
the following sections we use double-precision computation with
$\mbox{\em Tol}_g = 10^{-6}$ and $\mbox{\em Tol}_x = 10^{-5}$.

\subsection{Kicked double rotor map}\label{sec:drm}
The kicked double rotor map describes the dynamics of a
mechanical system known as the double rotor under the action
of a periodic kick~\cite{Romeiras92}. It is a four-dimensional map
defined by
\begin{equation}
  \left(\!\!\begin{array}{c}x_{n+1}\\y_{n+1}\end{array}\!\!\right) =
  \left(\!\!\begin{array}{l}My_{n} + x_{n}~(\mbox{mod}~2\pi)\\
  Ly_{n} + c\sin{x_{n+1}}\end{array}\!\!\right),
\end{equation}
where $x_{n}\in \mathbb{S}^2$ are the angle coordinates and
$y_{n}\in\mathbb{R}^2$ are the angular velocities after each kick.
Parameters $L$ and $M$ are constant $2\times 2$ matrices that depend
on the masses, lengths of rotor arms, and friction at the pivots,
while $c\in\mathbb{R}^2$ is a constant vector whose magnitude is
proportional to the kicking strength $f_0$.  In our numerical tests
we have used the same parameters as in~\cite{Romeiras92}, with the
kicking strength $f_{0} = 8.0$.

The following example illustrates the stabilising properties of the
transformations constructed on the basis of periodic orbits.  Let us
take a typical period-3 orbit point $x^\ast = (0.6767947,
5.8315697)$, $y^\ast = (0.9723920, -7.9998313)$ as a seed for
locating period-4 orbits.  The Jacobian matrix
$D\!f^3(x^\ast,y^\ast)$ of the seed has eigenvalues $\Lambda =
\mbox{diag}(206.48, -13.102, -0.000373, 0.000122)$. Therefore, based
on the scheme discussed in \S\ref{sec:ext}
Eqs.~(\ref{eq:gev}-\ref{eq:signs}), we can construct four
stabilising transformations $C$ corresponding to $(S_{11},S_{22})$
in Eq.~(\ref{eq:signs}) being equal to $(+,+)$, $(-,+)$, $(+,-)$ and
$(-,-)$.  Of the total of 2190 orbit points of period-4 (see Table
\ref{tab:drmp}), the transformations $C_1$, $C_2$, $C_3$, and $C_4$
stabilise \#$(1) = 532$, \#$(2)=544$, \#$(3)=474$, and \#$(4)=516$
orbit points, respectively, and these sets of orbits are almost
completely non-overlapping.  That is, the number of orbits
stabilised by both $C_1$ and $C_2$ is \#$(1\cap 2) = 2$.  Similarly,
\#$(1\cap 3) = 16$, \#$(1\cap 4) = 0$, \#$(2\cap 3) = 0$, \#$(2\cap
4) = 14$, and \#$(3\cap 4) = 0$. On the other hand, the number of
period-4 orbits stabilised by at least one of the four
transformations is \#$(1\cup 2\cup 3\cup 4) = 2034$.  This is a
typical picture for other seeds of period-3 as well as other
periods.

This example provides evidence for the validity of our approach to
constructing the stabilising transformations in high-dimensional
systems based on periodic orbits.  It also shows that, in the case
of the double rotor map, a single seed is sufficient for
constructing transformations that stabilise majority of the UPOs. Of
course, in order to locate the UPOs, we need to ensure that the
seeds are in the convergence basins of the stabilised periodic
orbits.  That is why we need to use more seeds to locate plausibly
all periodic orbits of a given period.  Still, because of the
enlarged basins of the stabilised orbits, the number of seeds is
much smaller than that required with iterative schemes that do not
use the stabilising transformations.

\begin{table}
\caption{Number $n(p)$ of prime period-$p$ UPOs, and the number
$N(p)$ of fixed points of $p$-times iterated map for the kicked
double rotor map.  The asterisk for $p=8$ indicates that this set of
orbits is not complete.  Parameters $\epsilon_\mathrm{max}(p)$ and
$d_\mathrm{min}(p)$ are defined in Eqs.~(\ref{eq:maxe}) and
(\ref{eq:mind}).}
\begin{center}\footnotesize
\begin{tabular}{|l|r|r|c|c|} \hline
$p$ & $n(p)$~~ & $N(p)$~~ & $\epsilon_\mathrm{max}(p)$ &
$d_\mathrm{min}(p)$\\\hline
1 &      12 &       12 & $1.0\cdot10^{-14}$ & $1.3\cdot10^{0}$\\
2 &      45 &      102 & $5.9\cdot10^{-14}$ & $3.4\cdot10^{-1}$\\
3 &     152 &      468 & $5.8\cdot10^{-13}$ & $6.2\cdot10^{-2}$\\
4 &     522 &     2190 & $2.7\cdot10^{-12}$ & $6.9\cdot10^{-3}$\\
5 &    2200 &  11\,012 & $2.6\cdot10^{-11}$ & $1.1\cdot10^{-3}$\\
6 &    9824 &  59\,502 & $1.6\cdot10^{-10}$ & $1.8\cdot10^{-4}$\\
7 & 46\,900 & 328\,312 & $9.7\cdot10^{-10}$ & $9.1\cdot10^{-5}$\\
8$^\ast\!\!$ & 229\,082 & 1\,834\,566 & $1.2\cdot10^{-8}$ &
$5.5\cdot10^{-5}$\\\hline
\end{tabular}
\end{center}
\label{tab:drmp}
\end{table}

Compared to the total of 384 matrices in ${\mathcal
C}_{\mathrm{SD}}$, we use only two or four transformations for each
seed, depending on the number of unstable directions of the seed
orbit points. Yet, the application of the detection scheme outlined
in \S\ref{sec:numres} allows us to locate plausibly all periodic
orbits of the double rotor map up to period 7.  Table \ref{tab:drmp}
also includes the number of detected period-8 orbits that were used
as seeds to complete the detection of period 7.

The confidence with which we claim to have plausibly complete sets
of periodic orbits for each period is enhanced by the symmetry
consideration. That is, since the double rotor map is invariant
under the change of variables $(x, y) \mapsto (2\pi - x, -y)$, a
necessary condition for the completeness of the set of orbits for
each period is that for any orbit point $(x^{*},y^{*})$ the set also
contains an orbit point $(2\pi - x^{*},-y^{*})$.  Even though this
condition was not used in the detection scheme, we find that the
detected sets of orbits (apart from period 8) satisfy this symmetry
condition.  Of course, this condition is not sufficient to prove the
completeness of the detected sets of UPOs, but, combined with the
exhaustive search procedure presented above, provides a strong
indication of the completeness.

\subsection{Coupled H\'{e}non maps}\label{sec:chm}
Another system we use to test the efficacy of our approach is
a six-dimensional system of three coupled H\'{e}non maps (CHM),
\begin{equation}\label{eq:coupled1}
  x^{j}_{n+1} = a - (\tilde{x}^{j}_{n})^{2} + bx^{j}_{n-1},
  \quad\mathrm{for}\quad j=1,2,3\,,
\end{equation}
where $a = 1.4$ and $b = 0.3$ are the standard parameter values of
the H\'{e}non map and the coupling is given by
\begin{equation}
  \tilde{x}^{j}_{n} = (1-\epsilon)x^{j}_{n} + \frac{1}{2}\epsilon(
   x^{j+1}_{n} + x^{j-1}_{n}),
\end{equation}
with $x^0_n \equiv x^3_n$ and $x^4_n \equiv x^1_n$.
We have chosen the coupling parameter $\epsilon = 0.15$.
Our choice of this system is motivated by the work
of Politi and Torcini~\cite{Politi92} in which they locate periodic
orbits in CHM for a small coupling parameter by extending the method
of Biham and Wenzel~\cite{Biham89}.  This makes the CHM an excellent
test system, since we can compare our results against those for the
Biham-Wenzel (BW) method. The BW method defines the following
artificial dynamics
\begin{equation}\label{eq:EBWflow}
\dot{x}_{n}^{j}(t) = (-1)^{s(n,j)}\{x^{j}_{n+1}(t) - a +[
                      \tilde{x}^{j}_{n}(t)]^{2} - bx^{j}_{n-1}(t)\},
\end{equation}
with $s(n,j)\in\{0,1\}$.  Given the boundary condition
$x_{p+1}^j = x_1^j$, the equilibrium states of Eq.~(\ref{eq:EBWflow})
are the period-$p$ orbits for the CHM.  The BW method is based on
the property that every equilibrium state of Eq.~(\ref{eq:EBWflow})
can be made stable by one of the $2^{3p}$ possible sequences of
$s(n,j)$ and, therefore, can be located by simply integrating
Eq.~(\ref{eq:EBWflow}) to convergence starting from the same initial
condition $x_n^j = 0.0$.  It is also found that,
for the vast majority of orbits, each orbit is stabilised by a
unique sequence of $s(n,j)$.

In order to reduce the computational effort Politi and Torcini
suggest reducing the search to only those sequences $s(n,j)$ which
are allowed in the uncoupled system, i.e. with $\epsilon = 0$.
This reduction is possible because the introduction of coupling has
the effect of pruning some of the orbits found in the uncoupled
H\'{e}non map without creating any new orbits.

\begin{table}
\caption{The number of prime UPOs for the system of three coupled
H\'{e}non maps (CHM) detected by three different methods: BW -- full
Biham-Wenzel, BW-r -- reduced Biham-Wenzel, ST -- our method based
on stabilising transformations, Max -- maximum number of detected
UPOs obtained from all three methods and the system symmetry. See
text for details.}
\begin{center}\footnotesize
\begin{tabular}{|l|r|r|r|r|c|c|} \hline
$p$ &  BW & BW-r &   ST &  Max & $\epsilon_\mathrm{max}(p)$ &
$d_\mathrm{min}(p)$\\ \hline
 1 &    8 &    8 &    8 &    8 & $1.3\cdot10^{-14}$ & $9.9\cdot10^{-1}$\\
 2 &   28 &   28 &   28 &   28 & $4.6\cdot10^{-14}$ & $5.2\cdot10^{-1}$\\
 3 &    0 &    0 &    0 &    0 &    -               &       -          \\
 4 &   34 &   34 &   40 &   40 & $2.7\cdot10^{-8}$  & $4.2\cdot10^{-2}$\\
 5 &    0 &    0 &    0 &    0 &    -               &       -          \\
 6 &   74 &   74 &   72 &   74 & $9.5\cdot10^{-10}$ & $8.6\cdot10^{-3}$\\
 7 &   28 &   28 &   28 &   28 & $1.0\cdot10^{-8}$  & $5.6\cdot10^{-3}$\\
 8 &  271 &  271 &  285 &  286 & $1.1\cdot10^{-6}$  & $5.5\cdot10^{-3}$\\
 9 &    - &   63 &   64 &   66 & $9.9\cdot10^{-7}$  & $2.6\cdot10^{-4}$\\
10 &    - &  565 &  563 &  568 & $1.3\cdot10^{-8}$  & $4.1\cdot10^{-4}$\\
11 &    - &  272 &  277 &  278 & $7.1\cdot10^{-9}$  & $5.4\cdot10^{-4}$\\
12 &    - & 1972 & 1999 & 1999 & $2.5\cdot10^{-6}$  & $4.3\cdot10^{-4}$\\
13$^\ast$& - & - & 1079 &   -  & $8.6\cdot10^{-8}$  & $4.0\cdot10^{-4}$\\
14$^\ast$& - & - & 6599 &   -  & $2.3\cdot10^{-6}$  & $3.5\cdot10^{-4}$\\
15$^\ast$& - & - & 5899 &   -  & $7.0\cdot10^{-6}$  &
$1.5\cdot10^{-4}$\\\hline
\end{tabular}
\end{center}
\label{tab:chm}
\end{table}

We have implemented the BW method with both the full search and the
reduced search (BW-r) up to as high a period as is computationally
feasible (see Table~\ref{tab:chm}).  In the case of the full search we
detect UPOs up to period 8 and in the case of the reduced search up
to period 12.  The seed $x_n^j = 0.0$ was used for all periods except
for period 4, where it was found that with this seed both BW and BW-r
located only 28 orbits.  We found a maximum of 34 orbits using
the seed $x_n^j = 0.5$.  It is possible that more orbits can be
found with different seeds for other periods as well, but we have not
investigated this.  The example of period 4 illustrates that, unlike
for a single H\'{e}non map, the Biham-Wenzel method fails to detect
all orbits from a single seed.

Even though our approach (labeled ``ST'' in Table~\ref{tab:chm}) is
general and does not rely on the special structure of the H\'{e}non
map, its efficiency far surpasses the full BW method and is
comparable to the reduced BW method.  Except for periods 6 and 10,
the ST method locates the same or larger number of
orbits.\footnote{The precise reason for the failure of the ST method
to detect all period 6 and 10 orbits needs further investigation.
We believe that the orbits that were not detected have
uncharacteristically small convergence basins with any of the
applied stabilising transformations.}

Unlike the double rotor map, the CHM possesses very few periodic
orbits for small $p$, particularly for odd values of $p$.
Therefore, we found that the direct application of the detection
strategy outlined at the beginning of \S\ref{sec:numres} would
not allow us to complete the detection of even period orbits.
Therefore, for even periods $p$ we also used $p+2$ as seeds and,
in case of period 12, a few remaining orbits were located with seeds
of period 15.  We did not attempt to locate a maximum possible number
of UPOs for $p > 12$.  The numbers of such orbits (labeled with
asterisks) are listed in Table~\ref{tab:chm} for completeness.

As with the double rotor map, we used the symmetry of the CHM to
test the completeness of the detected sets of orbits.  It is clear
from the definition of the CHM that all its UPOs are related by the
permutation symmetry (i.e., six permutations of indices $j$).  The
column labeled ``Max'' in Table~\ref{tab:chm} lists the maximum
number of UPOs that we were able to find using all three methods and
applying the permutation symmetry to find any UPOs that might have
been missed.  As can be seen in Table~\ref{tab:chm}, only a few
orbits remained undetected by the ST method.

Concluding this Section, we would like to point out that the high
efficiency of the proposed method is primarily due to the fact that
each stabilising transformation constructed based on the stability
properties of the seed orbit substantially increases the basins
of convergence of orbits stabilised by this transformation.  This
is apparent in a typical increase of the fraction of converged
seeds with the increasing value of parameter $\beta$ in
Eq.~(\ref{eq:itrDL}).  For example, when detecting period-10
orbits of CHM using period-12 orbits as seeds, the fraction of seeds
that converge to periodic orbits grows from 25-30\% for small $\beta$
(essentially the Newton-Raphson method) to about 70\% for the optimal
value of $\beta$.

\section{Discussion and Conclusions}
\label{sec:last} We have presented a new scheme for constructing
stabilising transformations which can be used to locate periodic
orbits in chaotic maps with the iterative scheme given by
Eq.~(\ref{eq:itrDL}).  The scheme is based on the understanding of
the relationship between the stabilising transformations and the
properties of eigenvalues and eigenvectors of the stability matrices
of the periodic orbits.  Of particular significance is the
observation that only the unstable eigenvalues are important for
determining the stabilising transformations.  Therefore, unlike the
original set of transformations proposed by Schmelcher and Diakonos,
which grows with the system dimensionality as $2^n n!$, our set has
the size of at most $2^k$, where $k$ is the maximum number of
unstable eigenvalues (i.e.~the maximum dimension of the unstable
manifold).  It is also apparent that, while the SD set contains a
large fraction of transformations that do not stabilise any UPOs of
a given system, all of our transformations stabilise a significant
subset of UPOs.  The dependence of the number of transformations on
the dimensionality of the unstable manifold rather than on the
system dimensionality is especially important in cases when we study
low-dimensional chaotic dynamics embedded in a high-dimensional
phase space.  This is often the case in systems obtained from
time-space discretisation of nonlinear partial differential
equations (e.g. the Kuramoto-Sivashinsky equation). Application of
the stabilising transformations approach to such high-dimensional
chaotic systems will be the subject of our future work.

The new transformations were tested on two systems: a kicked double
rotor map and three symmetrically coupled H\'{e}non maps. We aimed
to achieve a plausibly complete detection of periodic orbits of low
periods up to as high a period as was computationally feasible. In
both cases our algorithm was able to detect large numbers of UPOs
with high degree of certainty that the sets of UPOs for each period
were complete.  We have used the symmetry of the systems in order to
test the completeness of the detected sets.  On the other hand, when
the aim is to detect as many UPOs as possible without verifying the
completeness, the symmetry of the system could be used to increase
the efficiency of the detection of UPOs: once an orbit is detected,
additional orbits can be located by applying the symmetry
transformations.

One apparent drawback of the new scheme is that a small set of UPOs
needs to be available for the construction of the stabilising
transformation at the start of the detection process.  With the
systems studied so far, we had no problem detecting UPOs of low
period using the standard Newton-Raphson method by setting $\beta =
0$ in Eq.~(\ref{eq:itrDL}).  However, in systems where it is hard to
detect even a single periodic orbit, it would be useful to be able
to determine stabilising transformations without the knowledge of
UPOs. Since the stabilising transformations depend mostly on the
properties of the unstable subspace, and since the decomposition
into stable and unstable subspaces can be defined at any, not just
periodic, point on the chaotic set, it should be possible to
estimate such properties and construct the stabilising
transformations without the knowledge of the UPOs.  The
decomposition could be done, for example, in a process similar to
that used in the subspace iteration algorithm~\cite{Lust98}, and a
set of stabilising transformations, for example ${\mathcal
C}_{\mathrm{SD}}$, could then be applied only within the unstable
subspace.  The feasibility of such a construction will be the topic
for future investigation.

\section*{Acknowledgments}
The authors would like to thank Alexander Gorban for useful
discussion and numerous suggestions that helped improve the quality
of the manuscript.


\end{document}